\documentstyle[twocolumn,floats,aps,prb,epsf]{revtex}	
\title{Susceptibility inhomogeneity and non-Fermi liquid behavior \\ in 
nominally ordered $\bf UCu_4Pd$}
\author{D. E. MacLaughlin} 
\address{MS K764, Los Alamos National Laboratory, Los Alamos, New Mexico 87545 
\\ Department of Physics, University of California, Riverside, California 92521-0413
}
\author{R.~H. Heffner}
\address{MS K764, Los Alamos National Laboratory, Los Alamos, New Mexico 87545}
\author{G.~J. Nieuwenhuys}
\address{Kamerlingh Onnes Laboratory, Leiden University, 2300 RA Leiden, 
the Netherlands}
\author{G.~M. Luke, Y. Fudamoto, Y.~J. Uemura}
\address{Department of Physics, Columbia University, New York, NY 10027}
\author{R. Chau,\cite{LLL} M.~B. Maple}
\address{Department of Physics, University of California, San Diego, La Jolla, 
California 92093}
\author{B. Andraka}
\address{Department of Physics, University of Florida, Gainesville, Florida 
32611}
\author{\small(\today)}						

\address{\parbox{14cm}{\bigskip\rm\small			
Muon spin rotation experiments on a stoichiometric sample of the non-Fermi 
liquid (NFL) heavy-fermion compound~UCu$_4$Pd, in which recent neutron 
scattering experiments suggest an ordered structure, indicate that the U-ion 
susceptibility is strongly inhomogeneous at low temperatures. We argue that
this is due to residual disorder, which also dominates NFL behavior. The data 
yield a short correlation length ($\lesssim 1$ lattice spacing) and a rapid 
low-temperature U-moment relaxation rate ($\gtrsim 10^{12}\ {\rm s}^{-1}$), 
which constrain cluster-based models of NFL behavior.
\\[6pt] PACS numbers: 71.27.+a, 75.30.Mb, 76.60.Cq.}}	

\begin{document} \maketitle					

\thispagestyle{myheadings} \markright{{\small to be published in {\em Physical 
Review B (Rapid Communications)}}\hfill {\small LA-UR-98-1673\quad p.}\hspace{1mm}}

Many heavy-fermion alloys and compounds containing $f$ atoms have been found to 
exhibit thermodynamic and transport properties at low temperatures which are not 
in agreement with the conventional Landau Fermi-liquid predictions.\cite{ITP}\, 
Most such non-Fermi liquid (NFL) materials share two characteristics: proximity 
to a magnetic region of the appropriate phase diagram, and disorder due to chemical substitution. A 
number of mechanisms, some of which include one or both of these features, have 
been taken to be the basis for NFL behavior. This paper presents strong evidence 
for a disorder-driven NFL mechanism in nominally ordered UCu$_4$Pd.

Nuclear magnetic resonance\cite{BMLA95} (NMR) and muon spin 
rotation\cite{MBL96,BMAF96} ($\mu$SR) experiments in UCu$_{5-x}$Pd$_x$ alloys, 
which exhibit NFL behavior for $1 \lesssim x \lesssim 1.5$ 
(Ref.~\onlinecite{AS93}), revealed static linewidths which vary more rapidly 
than the bulk susceptibility with temperature and which become anomalously large 
at low temperatures. This is unambiguous evidence for strong inhomogeneity of 
the local magnetic susceptibility, which in turn suggests a large effect of 
disorder on the electronic structure of these alloys. This article describes 
$\mu$SR measurements in a new sample of UCu$_4$Pd, for which recent neutron 
diffraction experiments\cite{CMR98} have been interpreted as consistent with 
crystalline order. We find excellent agreement with the original $\mu$SR 
results, again indicating considerable susceptibility inhomogeneity. Either 
there is a residual level of disorder in this sample, which is possible given 
the uncertainty in the  neutron diffraction results, or UCu$_4$Pd exhibits a 
novel periodic spatial modulation of the local U-moment susceptibility in the 
absence of long-range spin order. We will argue that the former is most likely. 
We also find that the susceptibility inhomogeneity is characterized by a short 
correlation length (of the order of a lattice constant), and that the U moments 
fluctuate at a rate $\gtrsim 10^{12}\ {\rm s}^{-1}$. These latter results pose 
constraints for cluster-based theories of disorder-driven NFL behavior.

Hybridization between conduction-electrons and local moments gives rise to Kondo 
and other correlations in heavy-fermion systems.  Several treatments have 
appeared which ascribe NFL behavior to disorder in this hybridization. Bernal 
{\em et al.\/}\cite{BMLA95} proposed a simple phenomenological ``Kondo 
disorder'' model, in which the disorder results in a wide distribution of Kondo 
temperatures~$T_K$\@.  At a given temperature~$T$ local moments with $T_K < T$ 
are uncompensated and give rise to the NFL behavior. The wide spread of Kondo 
temperatures gives rise to a temperature-dependent spread~$\delta\chi$ of 
local-moment susceptibilities~$\chi(T,T_K)$. A similar picture was proposed 
independently by Matsuhira {\em et al.\/}\cite{MSA95} in connection with their 
experiments on randomly diluted CeRu$_2$Si$_2$. Miranda {\em et 
al.\/}\cite{MDK96} put this model on a firmer theoretical foundation, and 
pointed out that it was capable of explaining the ubiquitous linear temperature 
dependence of the low-temperature electrical resistivity. 

Recently Castro Neto {\em et al.\/}\cite{CNCJ98} have extended the simple 
single-impurity Kondo disorder model to take into account the 
Ruderman-Kittel-Kasuya-Yosida (RKKY) coupling between local moments. They 
concluded that the combination of disorder and a critical point associated with 
cooperative behavior of these local moments give rise to a Griffiths phase of 
correlated clusters associated with a nonuniversal scaling exponent~$\lambda$. 
This theory is in good agreement with specific heat and susceptibility data for 
a number of NFL materials.\cite{dACD98}\, 

Disorder is an essential ingredient for these NFL mechanisms. Some NFL compounds 
may crystallize in ordered lattice structures, however, in which case there is 
no reason to suspect a disorder-driven mechanism for NFL behavior if the atomic 
order is perfect. UCu$_4$Pd is a candidate for such an ordered compound.

\begin{figure}[t]
\epsfxsize=3.4in \epsfbox{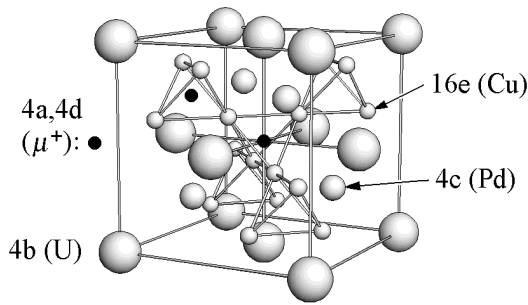}			
\caption{Unit cell of UCu$_4$Pd assuming structural order, with atom sites 
indicated in Wyckoff notation. Muon stopping sites in the end compound~UCu$_5$ 
(Ref.~\protect\onlinecite{BOGS86}) are also shown.} 
\label{fig:unitcell}
\end{figure}

UCu$_{5-x}$Pd$_x$, $0 \le x \lesssim 2.5$, crystallizes in the fcc AuBe$_5$ 
structure (space group~$F\overline{4}3m$). The end compound~UCu$_5$ possesses 
two crystallographically inequivalent copper sites in the ratio $4 : 1$ at the 
$16e$ and $4c$ positions (Wy\-ckoff notation). Thus stoichiometric UCu$_4$Pd 
could order as shown in Fig.~\ref{fig:unitcell}. 
The concentration dependence of the lattice parameter exhibits a break at $x 
\approx 1$ (Ref.~\onlinecite{AS93}), which suggests preferential Pd occupation 
of 4$c$ sites but does not determine the degree of order.

Recently Chau {\em et al.\/}\cite{CMR98} reported elastic neutron diffraction 
measurements on members of the UCu$_{5-x}$Pd$_x$ series. From Rietveld 
refinements of the data they find that indeed Pd atoms preferentially occupy 
$4c$ sites for $x < 1$, and a mixture of Cu and Pd atoms occupy $16e$ sites for 
$x > 1$. For $x = 1$ ``\dots there is no evidence for Pd/Cu 
disorder\dots''.\cite{CMR98}\, The data do not rule out the possibility of 
site interchange between Pd and Cu sites, at the level of $\sim$4.3\% occupation 
of $16e$ (Cu) sites by Pd atoms [and therefore $\sim$17\% occupation of $4c$ 
(Pd) sites by Cu atoms]. A minimum in the variation with $x$ of the residual 
resistance ratio~${\rm(RRR)} = R(300\ {\rm K})/R(1\ {\rm K})$ was also observed 
near $x = 1$, indicating that the electronic structure undergoes a transition 
near this Pd concentration. 

In view of these results we have carried out $\mu$SR studies on a portion of the 
same powder sample of UCu$_4$Pd used for the neutron diffraction 
measurements.\cite{CMR98}\, The data reproduce earlier results on a different 
UCu$_4$Pd sample\cite{BMAF96} extremely well, exhibiting the rapid increase of 
linewidth with decreasing temperature and strong inhomogeneity in the 
low-temperature susceptibility (distribution width/mean ${\rm susceptibility} 
\gtrsim 2$) predicted by disorder-driven theories of NFL 
behavior.\cite{BMLA95,MDK96,CNCJ98}\, 

Positive muons ($\mu^+$) from the M20 beam line at \mbox{TRIUMF}, Vancouver, 
Canada, were stopped in the sample and subsequently precessed in the sum of the 
applied field (if any) and the local internal field at the $\mu^+$ site. As in 
previous studies,\cite{BMAF96} zero-field $\mu$SR linewidths showed no evidence 
for spin freezing or static magnetism greater than ${\sim}0.01\ \mu_B$/U atom 
down to $\sim$2\ K\@. 

\begin{figure}[t]
\epsfxsize=3.4in \epsfbox{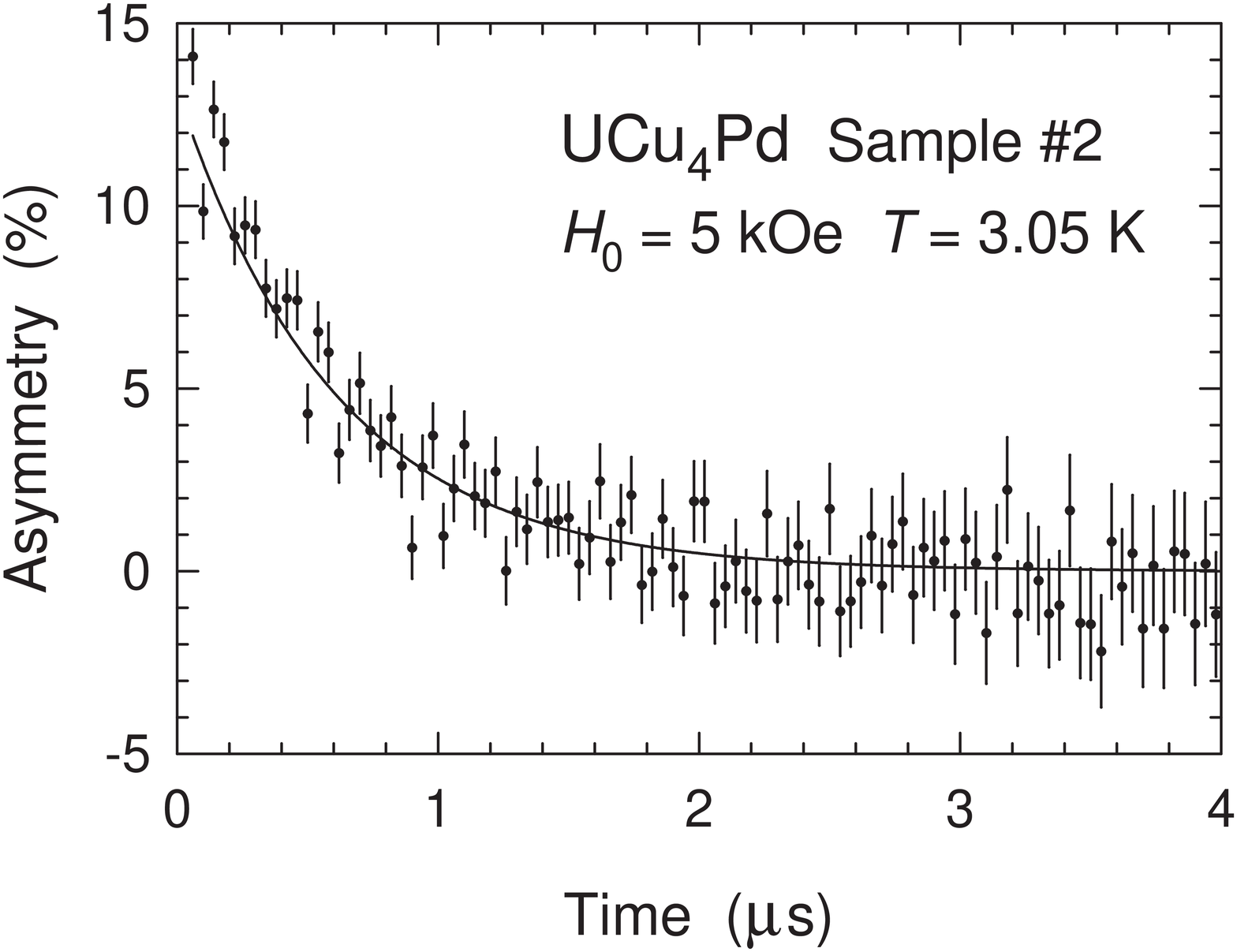}
\caption{Representative TF-$\mu$SR asymmetry relaxation function (after 
transformation to a reference frame rotating with the $\mu^+$ precession 
frequency) in UCu$_4$Pd sample (Sample~\#2) used in neutron diffraction studies 
(Ref.~\protect\onlinecite{CMR98}). Curve: best-fit exponential.}
\label{fig:histogram}
\end{figure}
The muon relaxation rate~$1/T_1$ in low ($\sim$100\ Oe) fields applied parallel 
to the muon spin, which is due to U-moment thermal fluctuations,\cite{Sche85} 
was found to be unobservably slow (${\lesssim}10^{4}\ {\rm s}^{-1}$) at all 
temperatures. For the present case of coupling between U and $\mu^+$ magnetic 
moments via a fluctuating dipolar field~$B_{\rm dip}$ 
(Ref.~\onlinecite{BMAF96}), the U-moment fluctuation rate~$\nu$ is given 
by\cite{Sche85} $1/T_1 = 2(\gamma_\mu B_{\rm dip})^2/\nu$, where $\gamma_\mu$ is 
the $\mu^+$ gyromagnetic ratio. We find a lower limit on $\nu$ of ${\sim}2\times 
10^{12}\ {\rm s}^{-1}$.

$\mu$SR spectra were obtained for applied fields~$H_0$ between 2 and 20\ kOe 
transverse to the $\mu^+$ spin direction (TF-$\mu$SR) over the temperature 
range~3--300\ K\@. A typical TF-$\mu$SR relaxation function is shown in 
Fig.~\ref{fig:histogram}, with the oscillatory factor removed by transforming to 
a reference frame rotating at the $\mu^+$ precession frequency.  It can be seen 
that the relaxation is exponential to within uncertainties. In spite of this the 
relaxation is not due to lifetime broadening, as evidenced by the negligible 
dynamic $\mu^+$ relaxation rate discussed above. Instead, the exponential 
relaxation indicates a Lorentzian distribution of $\mu^+$ Larmor frequencies. 

\begin{figure}[t]
\epsfxsize=3.4in \epsfbox{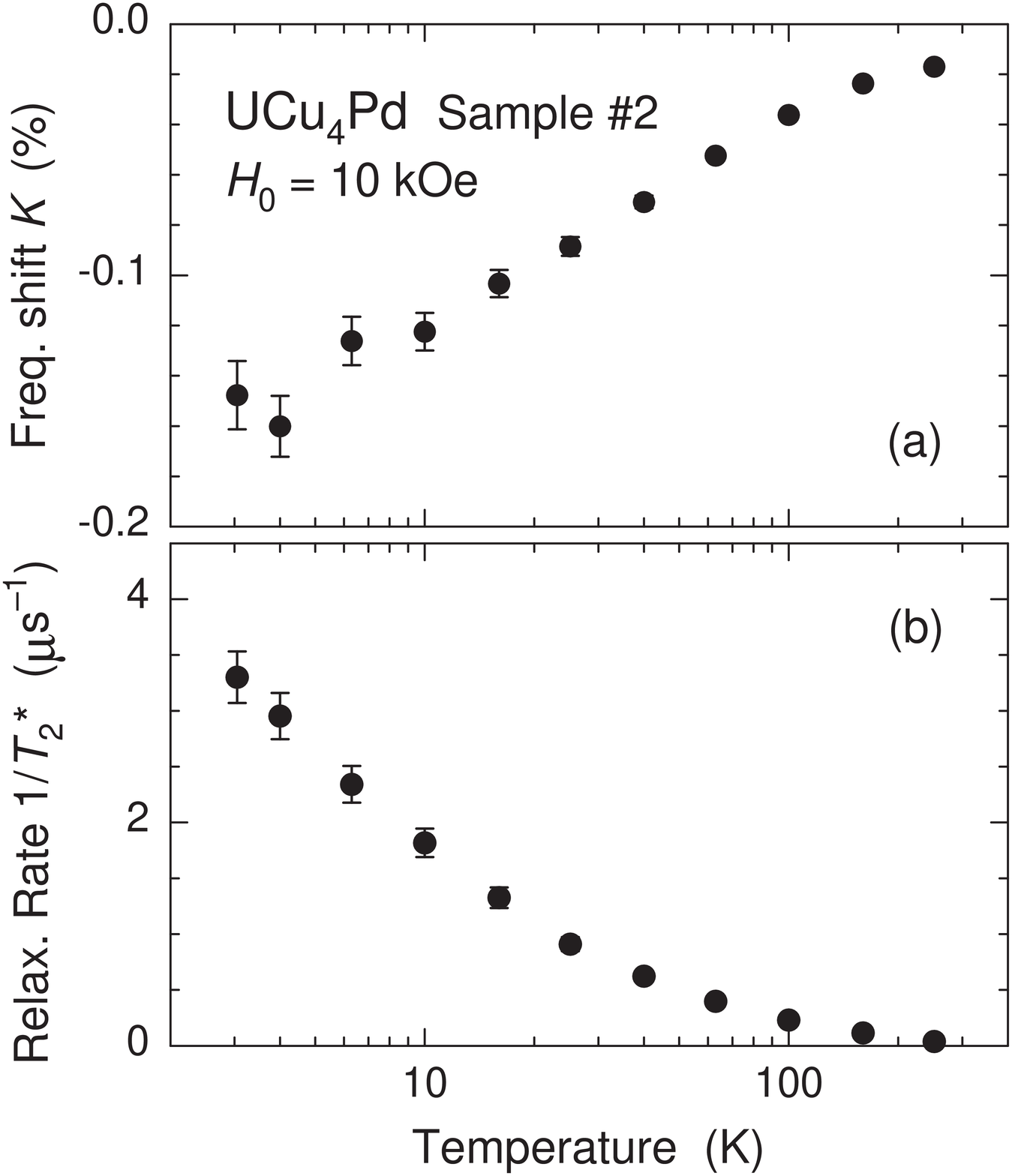}			
\caption{Temperature dependence of $\mu^+$ (a)~frequency shift $K$ and 
(b)~relaxation rate $1/T_2^{\scriptstyle*}$ in UCu$_4$Pd.}
\label{fig:tempdep}
\end{figure}

Frequency shifts~$K$ and exponential relaxation rates~$1/T_2^{\textstyle*}$ were 
obtained from fits to the relaxation function 
\begin{equation} 
G(t) = Ae^{-t/T_2^{\scriptstyle*}} \cos\left[\gamma_\mu H_0(1 
+ K) t + \phi\right] 
\end{equation}
to $\mu^+$ time-differential relaxation data. Here $A$ is the initial muon decay 
asymmetry, $H_0$ is the applied transverse field, and $\phi$ is the phase of the 
initial $\mu^+$ spin orientation. The temperature dependence of $K$ (not corrected 
for Lorentz and demagnetizing fields) and $1/T_2^{\textstyle*}$ at $H_0 = 10$\ kOe 
are shown in Fig.~\ref{fig:tempdep}. 
Both these quantities increase considerably in magnitude with decreasing 
temperature. In the case of $K$ this is expected, as both microscopic and 
macroscopic (Lorentz and demagnetization) contributions to $K$ should be 
proportional to the bulk susceptibility.\cite{BMAF96}\, At 3 K no significant 
field dependence of $K$ or of the half-width~$\delta K = 1/(\gamma_\mu 
H_0T_2^{\textstyle*})$ of the frequency shift distribution was observed for 
$H_0 < 20$\ kOe. 

Estimates of the relative spread~$\delta\chi/\chi$ were obtained using the 
relation
\begin{equation} 
\frac{\delta\chi}{\chi} = \frac{\delta K}{a^{\textstyle*}\chi} \,, 
\label{eq:dchionchi}
\end{equation}
where $a^{\textstyle*}$ is an effective hyperfine coupling constant between U 
moments and $\mu$ spins.\cite{MBL96,BMAF96}\, Assuming randomly-disordered 
inhomogeneity of the U-moment static susceptibility, $a^{\textstyle*}$ depends 
on the correlation length $\xi_\chi$ which describes the spatial variation of 
the inhomo\-geneity.\cite{MBL96}\, It was found previously from comparison of 
TF-$\mu$SR and NMR linewidths\cite{MBL96,BMAF96} that $\xi_\chi$ is of the order 
of a lattice parameter or smaller in UCu$_{5-x}$Pd$_x$, $x = 1.0$ and 1.5. This 
can be seen qualitatively as follows:\cite{MBL96} if $\xi_\chi$ were long ($\gg 
\mbox{a lattice parameter}$) the susceptibility in the vicinity of a given 
$\mu^+$ site would be nearly uniform. Then the dipolar $\mu^+$ hyperfine field 
would nearly vanish by symmetry at the $4a$ and $4d$ $\mu^+$ sites 
found\cite{BOGS86} in the end compound~UCu$_5$ (cf.\ Fig.~\ref{fig:unitcell}), 
which are octahedrally and tetrahedrally coordinated by U ions.  For long 
$\xi_\chi$, therefore, the distributed $\mu^+$ shifts and hence the linewidth 
would be small, contrary to observation. The vanishing dipolar field also rules 
out any contribution to the powder-pattern linewidth due to anisotropy of the 
shift.

\begin{figure}[t]
\epsfxsize=3.4in \epsfbox{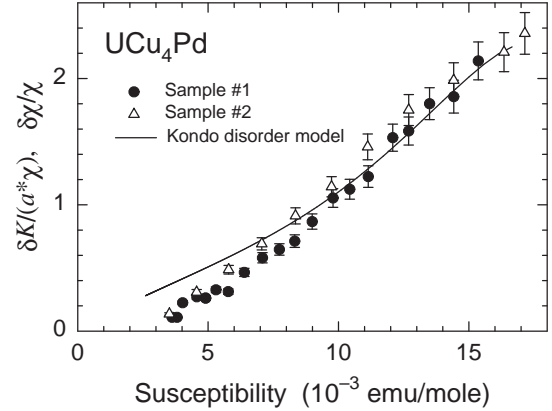}				
\caption{Dependence of $\delta K/(a^{\textstyle*}\chi)$ on bulk 
susceptibility~$\chi$, with temperature an implicit parameter, in UCu$_4$Pd. 
Symbols are defined in the text. Sample~\#1: previously-studied sample 
(Ref.~\protect\onlinecite{BMAF96}). Sample~\#2: present sample, used in neutron 
diffraction studies (Ref.~\protect\onlinecite{CMR98}). Curve: $\delta\chi/\chi$ 
from the Kondo disorder theory (Refs.~\protect\onlinecite{MBL96} and 
\protect\onlinecite{BMAF96}).} 
\label{fig:dchionchi}
\end{figure}

Figure~\ref{fig:dchionchi} plots $\delta K/(a^{\textstyle*}\chi)$ vs.\ the 
uniform bulk susceptibility, with temperature an implicit parameter. 
The data points were obtained from the $\mu$SR data and Eq.~(\ref{eq:dchionchi}) 
using the dipolar-coupling value~$\overline{a^{\textstyle*}} = 0.096\ 
{\rm mol}\,{\rm emu}^{-1}$ averaged over the two muon sites\cite{okay} and 
calculated assuming short-range correlation.\cite{MBL96}\, Data are shown for the 
previously-studied\cite{BMAF96} sample (Sample~\#1) (reanalyzed for this 
comparison) and the present sample (Sample~\#2). It can be seen that the 
agreement between the two data sets is very good.  

Also shown in Fig.~\ref{fig:dchionchi} for comparison is $\delta\chi/\chi$ from 
the Kondo disorder model.\cite{MBL96}\, The parameters defining the distribution 
function~$P(T_K)$ were obtained from fits to the field and temperature 
dependence of the bulk susceptibility, and $\delta\chi/\chi$ was then calculated 
with no further adjustable parameters. Comparison with the disorder-driven 
Griffiths-phase theory (not shown), made in a similar way, is also satisfactory 
(cf.\ Ref.~\onlinecite{CNCJ98}). Independently of agreement with a specific 
model, it is important to note that if the U-ion susceptibility were uniform 
(as in any NFL theory without disorder) $\delta K/(a^{\textstyle*}\chi)$ would be 
constant;\cite{MBL96} its strong dependence on $\chi$ and large low-temperature 
(large-$\chi$) value (Fig.~\ref{fig:dchionchi}) are unambiguous evidence for 
susceptibility inhomogeneity.

It is conceivable that disorder plays no role in the susceptibility 
inhomogeneity, which then must have an ordered (periodic) spatial distribution. 
Such a periodically modulated susceptibility cannot be due to long-range spin 
ordering, since as discussed above zero-field $\mu$SR shows that there is no 
static magnetism. We know of nothing to rule out a modulated susceptibility, but 
to our knowledge such behavior has never been observed without a corresponding 
modulated static moment (antiferromagnetism or a spin density wave). 

Thus residual disorder appears to be the most likely explanation of these results. 
The reproducibility of $\delta K/(a^{\textstyle*}\chi)$ between Sample~\#1 and 
Sample~\#2 (Fig.~\ref{fig:dchionchi}) is striking, however. The concentration of
lattice defects in an arc-melted sample depends on conditions such as vapor 
pressures, cooling rate, etc., which are not likely to be reproduced exactly in 
two different laboratories. We speculate that the two samples may not have 
the same lattice defect concentration, in which case the inhomogeneity is 
remarkably insensitive to the amount of disorder. We also note that the 
low-temperature values of $\delta K/(a^{\textstyle*}\chi)$ in UCu$_4$Pd and 
UCu$_{3.5}$Pd$_{1.5}$ (which is definitely a disordered alloy) are also very 
similar.\cite{BMAF96}\, These results raise the question of whether theories of 
NFL behavior which are {\em not\/} based on disorder can ever be verified in 
real compounds, which will always possess some level of defects. 
X-ray-absorption-fine-structure (EXAFS) and neutron pair-distribution-function 
(PDF) studies are under way to determine the amount of structural disorder in 
UCu$_4$Pd. 

These results also provide important constraints on ``cluster-based'' mechanisms 
for NFL behavior (e.g., Ref.~\onlinecite{CNCJ98}). First, the short correlation 
length discussed above seems to limit the number of $f$ ions in the clusters to 
the order of a few. Second, the absence of observable $\mu^+$ dynamic relaxation 
sets a lower limit of ${\sim}2 \times 10^{12}\ {\rm s}^{-1}$ on the fluctuation 
rate of spins in clusters. Any cluster-based mechanism must therefore be 
consistent with the small size and rapid fluctuation rate of the clusters.

In the absence of evidence for a periodically modulated susceptibility, we 
conclude that disorder-driven mechanisms are able to explain NFL behavior in 
UCu$_4$Pd. Further experimental work is needed to decide between various such 
mechanisms, and future NMR and $\mu$SR experiments will be necessary to characterize 
disorder-driven NFL behavior in other nominally ordered compounds. 

\medskip We are grateful to S.~R. Dunsiger, M. Good, B. Hitti, R.~F. Kiefl, and 
S.~R. Kreitzman for assistance during the experiments, and to M.~C. Aronson, 
W.~P. Beyermann, C.~H. Booth, A.~H. Castro Neto, V. Dobrosavljevi\'c, J.~M. 
Lawrence, A.~J. Millis, R. Osborn, and J.~D. Thompson for illuminating 
discussions. This research was supported in part by the U.S. National Science 
Foundation, Grants DMR-9418991 (U.C. Riverside), DMR-9510454 (Columbia), 
DMR-9705454 (U.C. San Diego), and DMR-9400755 (U. Florida), and by the U.C. 
Riverside Academic Senate Committee on Research, the Japanese agency NEDO 
(Columbia), and the Netherlands agencies FOM and NWO (Leiden), and was carried 
out in part under the auspices of the U.S. DOE (Los Alamos).


\begin{references}
\bibitem{okay} the values of $a^{\textstyle*}$ for the two sites differ from this 
average by $\sim$10\%, which does not affect the conclusions of this paper.
\end{references}
\end{document}